\documentclass[utf8,usenames,dvipsnames]{FrontiersinHarvard}
\definecolor{mysectioncolor}{RGB}{226, 124, 70}

\usepackage{tgpagella}
\usepackage{titlesec}
\titleformat{\section}
{\color{mysectioncolor}\fontfamily{qag}\selectfont\Large}
{\thesection}{1em}{}
\titleformat{\subsection}
{\color{black}\fontfamily{qag}\selectfont\large}
{\thesubsection}{1em}{}
\usepackage{url,hyperref,lineno,microtype,subcaption}
\usepackage{graphics}
\usepackage{upgreek}
\usepackage{xspace}
\usepackage{cleveref}
\usepackage{siunitx}

\hypersetup{
    colorlinks=true,           
    linkcolor=mysectioncolor,  
    citecolor=mysectioncolor,  
    urlcolor=mysectioncolor    
}

\makeatletter
\renewcommand{\normalsize}{\fontsize{8}{10}\selectfont}
\renewcommand{\large}{\fontsize{11}{10}\selectfont}
\renewcommand{\Large}{\fontsize{13}{10}\selectfont}

\makeatother
\begin{document}
\newcommand{\soneuv}{\ensuremath{\mathrm{S1}_\mathrm{UV}}\xspace}
\newcommand{\stwouv}{\ensuremath{\mathrm{S2}_\mathrm{UV}}\xspace}

\newcommand*{\todo}{\textcolor{red}}

\onecolumn

\def\keyFont{\fontsize{8}{11}\helveticabold }
\def\firstAuthorLast{Hammann {et~al.}}
\def\Authors{Robert~Hammann\,$^{1\star}$, Kai~B\"ose\,$^{1\star}$, Steffen~Form\,$^{1}$, Luisa H\"otzsch\,$^{1}$\footnote{Present address: Physik-Institut, University of Z\"urich, Switzerland}~ and Teresa~Marrod\'an~Undagoitia\,$^{1}$}
\def\Address{$^{1}$Max-Planck-Institut f\"ur Kernphysik, Heidelberg, Germany} 
\def\corrAuthor{Robert Hammann}
\def\corrEmail{robert.hammann@mpi-hd.mpg.de}
\extraAuth{Kai B\"ose \\ kai.boese@mpi-hd.mpg.de}
\author[\firstAuthorLast ]{\Authors} 
\address{} 
\correspondance{} 
\title[Xenon infrared emission]{Operation of a dual-phase xenon detector with wavelength sensitivity from ultraviolet to infrared}

\firstpage{1}

\maketitle

\begin{abstract}
\begingroup
\fontsize{9}{13}\selectfont

\section{}
\noindent
Xenon, in both its gaseous and liquid phase, offers excellent scintillation and ionization properties, making it an ideal target medium for rare event searches.
We report on measurements performed with a dual-phase xenon time projection chamber sensitive to wavelengths from 170\,nm to 1700\,nm.
In addition to the well-established ultraviolet (UV) scintillation, we observe coincident signals in a photomultiplier tube sensitive to infrared (IR) light, associated with both prompt scintillation in the liquid and electroluminescence in the gas.
We study the time response of the IR signals and their dependence on the applied amplification field in the gas.
Our findings support the observation of IR emission from electroluminescence and reveal a time response distinct from that previously reported for $\alpha$-particles in gas.
The results suggest that IR scintillation could provide enhanced signal identification and background rejection in future xenon-based detectors.
 
\endgroup
\tiny
 \keyFont{ \section{Keywords:} liquid xenon, scintillation, infrared radiation, dark matter, noble gas detectors, time projection chamber} 
\end{abstract}

\twocolumn
\section{Introduction} 
\label{sec:intro}

Xenon is widely used as a detector medium in various rare event searches, including dark matter direct detection and neutrinoless double beta decay.
Its application spans both liquid xenon\,(\cite{XENON:2025_wimp,LZ:2025_wimp}) and high-pressure gaseous xenon~(\cite{NEXT:2025yqw, Bouet:2024neu, PandaX:2017}).
The success of xenon in these fields is largely attributed to its high stopping power and strong vacuum ultraviolet (VUV) scintillation, which enables excellent energy resolution.

In addition to the well-studied VUV scintillation, xenon is also known to emit radiation in the infrared (IR) region. However, this IR emission remains relatively unexplored, particularly with regard to its potential implications for detector performance and background discrimination.

A first systematic investigation of IR scintillation in noble gases was conducted by~\cite{Lindblom:1988_nir}.
Around the year 2000, the first studies of xenon’s IR continuum emission identified a broad scintillation peak centered around 1.3\,$\upmu$m\,(\cite{Borghesani:2001xx}), consistent with earlier predictions from potential curve estimates\,(\cite{Mulliken:1970rs}). 
Initial measurements were conducted in gaseous xenon\,(\cite{Carungo:1998xx}), where the IR light yield was found to be comparable in magnitude to that of the UV scintillation\,(\cite{Belogurov:2000}). A red shift of the emission peak with increasing pressure was also observed and later modeled\,(\cite{Borghesani_2007,Borghesani:2007b,Borghesani:2025xx}).
Above an electric field of about 500\,V/cm in the gas, the IR signal was found to increase roughly in proportion to the ionization charge, consistent with a signal from electroluminescence\,(\cite{Belogurov:2000}).
IR scintillation has additionally been detected in liquid xenon\,(\cite{Bressi:2000nim}), though with a light yield estimated to be two orders of magnitude lower than in gas.
Furthermore, there are indications that the emission spectrum in the liquid is significantly shifted toward shorter wavelengths\,(\cite{Bressi:2001}).

To better understand xenon’s IR scintillation properties, we initiated a dedicated R\&D program.
As a first step, we operated a room-temperature setup using gaseous xenon near atmospheric pressure.
Using this setup, we studied the time response and light yield of IR scintillation following $\alpha$-particle irradiation, under varying pressure and gas purity. We observed that, at approximately 1 bar, the light yield is about one-quarter of that in the UV.
We found that the time response is well described by a sum of two exponential decays: a fast decay with a time constant of $\mathcal{O}(\mathrm{ns})$, contributing less than 1\,\% of the total yield, and a slower decay with a time constant of $\mathcal{O}(\mathrm{\upmu s})$\,(\cite{Piotter:2023qli,Hammann:2024qyy}).
We observed a linear increase in light yield with pressure between 500 and 1100\,mbar, along with a decrease in yield as purity decreased.
These measurements were performed using a PMT sensitive to wavelengths between about 950 and 1700\,nm.

In this paper, we report on the operation of a dual-phase xenon time projection chamber (TPC) instrumented with a UV- and an IR-sensitive PMT,  enabling simultaneous measurement of both scintillation components. To our knowledge, this is the first measurement in a TPC with wavelength sensitivities from 170\,nm to 1700\,nm.

The document is organized as follows: in \cref{sec:setup}, we describe the experimental setup. \Cref{sec:data} covers the calibration source, the data acquisition (DAQ), the selection criteria applied to the data, and the analysis procedure. The results are presented in \cref{sec:results} and discussed in \cref{sec:discussion}.

\section{Experimental setup}   
\label{sec:setup}
The measurements presented here were carried out using a small-scale liquid xenon TPC called HeXe (short for Heidelberg Xenon, operated at the Max-Planck-Institut f\"ur Kernphysik).
The detector consists of a $\sim350$\,g liquid xenon target with a thin layer of gaseous xenon above it serving as an amplification region, and is instrumented with two PMTs.
Energy depositions in a liquid xenon target induce both excitation and ionization of the medium. The excited states undergo relaxation processes that lead to the emission of scintillation light. We refer to the UV component of this prompt emission as \soneuv. Electrons liberated through ionization are drifted upward by an electric drift field to the liquid-gas interface. A stronger electric field is employed to extract the electrons from the liquid into the gas phase.
There, the high field accelerates the electrons, enabling them to excite xenon atoms and produce secondary scintillation light through electroluminescence\,(\cite{Lansiart:1976}).
We denote the resulting UV emission as \stwouv.

\subsection{The HeXe TPC}  
\label{subsec:tpc}
\begin{figure}[h]
	\centering
	\includegraphics[width=\columnwidth]{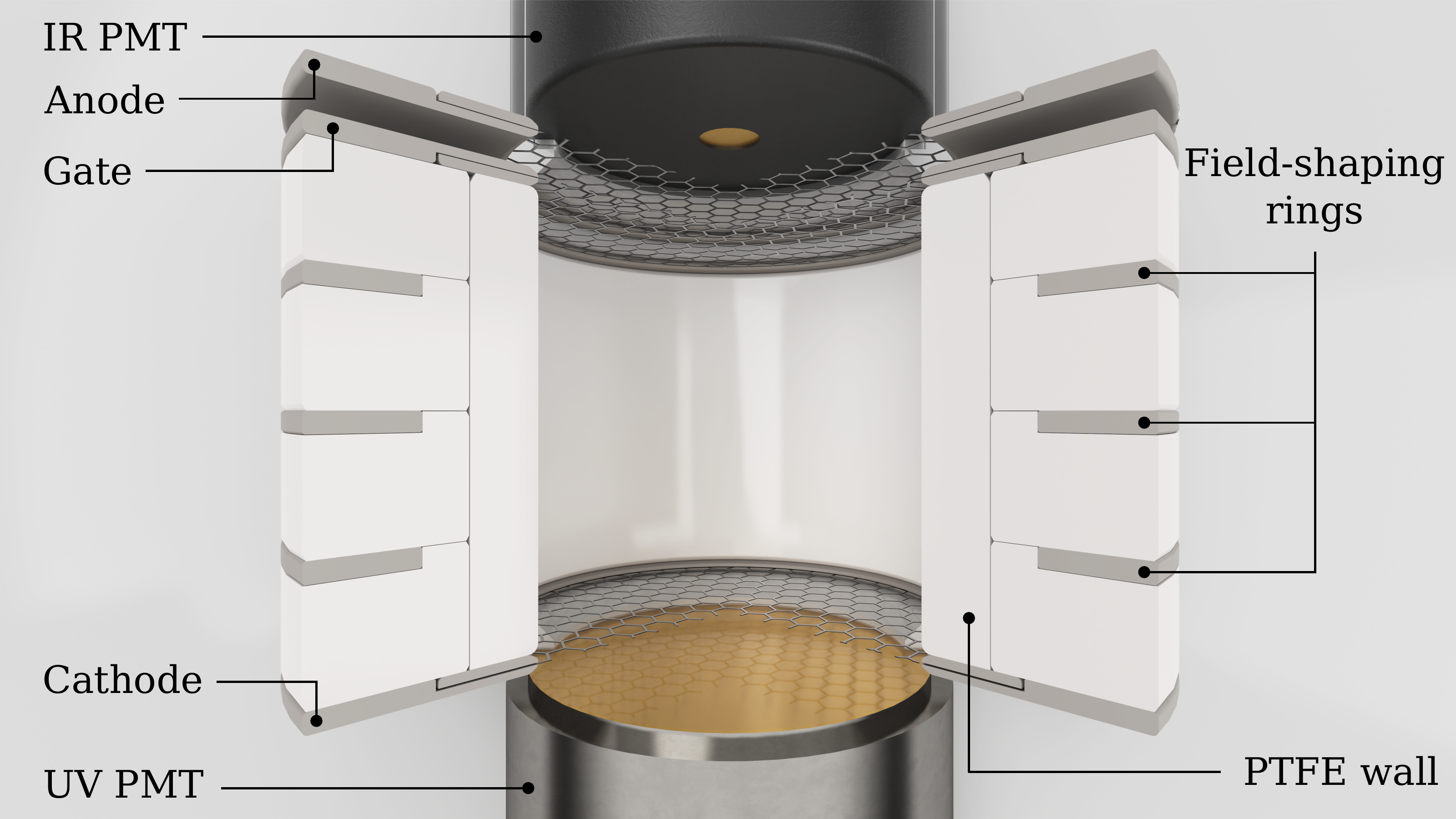}
	\caption{Schematics of the HeXe TPC with the IR-sensitive PMT at the top (aperture opening at the center) and the UV-sensitive PMT at the bottom. The cathode (bottom), gate, and anode (top) electrodes are visible. The 5 cm-high cylindrical TPC is enclosed by a wall made of PTFE. During operation, the liquid xenon level is maintained between the gate and anode.}
	\label{fig:TPC_ir}
	\end{figure}

The HeXe TPC and the auxiliary systems for its operation are detailed in~(\cite{Jorg:2021hzu}).
\Cref{fig:TPC_ir} illustrates the TPC configuration used in this work, where the upper PMT has been replaced with an IR-sensitive one, which is described in \cref{subsec:pmt}. The TPC’s electric fields are established by three grid electrodes: the cathode at the bottom, the gate just below the liquid surface, and the anode positioned right above the liquid. 
Field-shaping rings ensure a homogeneous electric field across the sensitive volume. Polytetrafluoroethylene (PTFE) is used both as a reflector and as a filler material between the cryostat and the target.

The drift field is kept constant for all measurements at about 400\,V/cm by setting the cathode to $-1950$\,V and the gate to $-25$\,V, spaced 5\,cm apart. This results in a maximum drift time of about 30\,$\upmu$s. The distance between gate and anode is 5\,mm. While the liquid level is typically set to 2.5\,mm above the gate, midway between gate and anode, it was ($4.1 \pm 0.5$)\,mm in this run.
The anode was set to $3750$\,V, resulting in an amplification field of about 12\,kV/cm in the gas.
During operation, the temperature was held constant at about $-95$\,$^\circ$C and the pressure in the gas phase was maintained at $2.5$\,bar.

\subsection{The infrared-sensitive PMT}  
\label{subsec:pmt}

\begin{figure*}[ht]
	\centering
	\includegraphics[width=\textwidth]{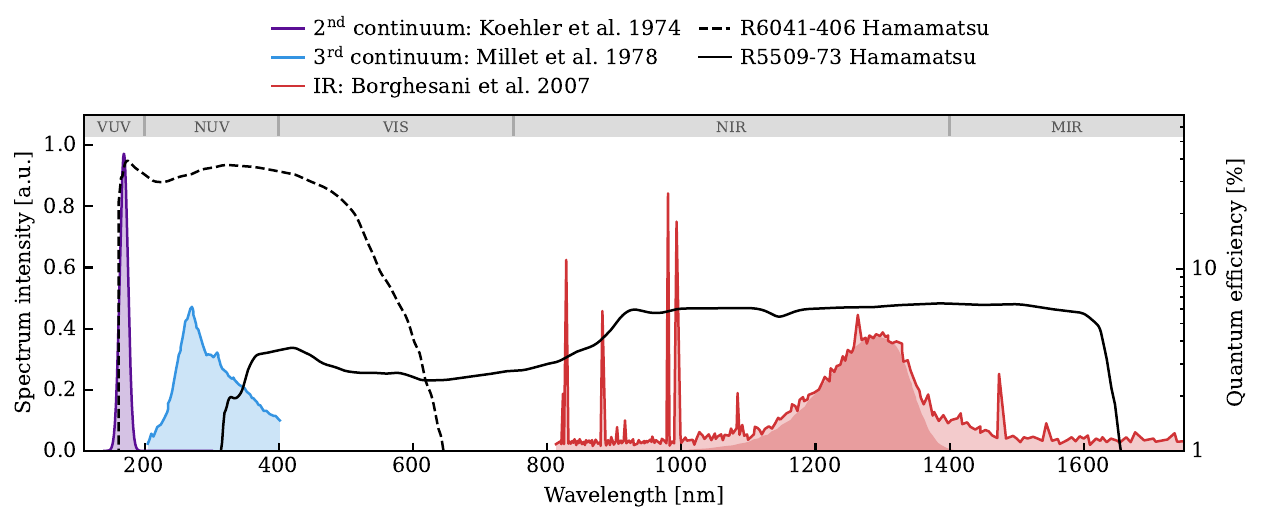}
	\caption{Nominal quantum efficiencies of the PMTs used in this setup: R5509-73 (IR-sensitive, solid black) and R6041-406 (UV-sensitive, dashed black), shown alongside the emission spectra of gaseous xenon at about atmospheric pressure. The UV emission from the second continuum is shown in purple (\cite{Koehler:1974}), the third continuum emission is shown in blue (\cite{Millet:1978}), and the IR emission in red (\cite{Borghesani_2007}). For the IR emission, the continuum region is indicated by a darker-shaded area. Note that the relative intensities among spectra are arbitrary. For reference, VUV, near-UV (NUV), visible (VIS), near-IR (NIR), and mid-IR (MIR) wavelength bands are indicated. Note that in the liquid phase, clear evidence of emission exists only for the VUV component.}
	\label{fig:spectra_and_qe}
	\end{figure*}

To record the IR scintillation light, we employ a 2-inch Hamamatsu R5509-73 PMT.
It features a borosilicate entrance window with an opening of 8\,mm diameter (see top PMT in \cref{fig:TPC_ir}). Its reflexion-type photocathode, measuring $3\,\textrm{mm}\times 8\,\textrm{mm}$ and made 
out of InP/InGaAs, is located 2\,cm behind the window.
The PMT is sensitive to wavelengths between 300\,nm and 1.65\,$\upmu$m (see \cref{fig:spectra_and_qe}), with a nominal quantum efficiency (QE) of about 6\% at the expected xenon IR scintillation wavelength of 1.3\,$\upmu$m (for gaseous xenon at 1\,bar). The PMT was operated at $-1700$ V for all measurements, corresponding to a gain of $3\times 10^6$, as stated by the manufacturer.  

To achieve optimal QE and dark count rate, the photocathode of the specific PMT used in this work must be biased at $-4.75$\,V to lower the conduction band barrier\,(\cite{Hamamatsu:2017}). 
This was implemented by stepping down the supply voltage using a Zener diode in series with a potentiometer, allowing the voltage to be adjusted. The potentiometer was placed outside the cold region.
During operation in the TPC, however, the bias voltage of the PMT occasionally exhibited discrete jumps to other constant values and was generally lower than designed (at $\sim\,-1$\,V).
The cause is still under investigation, but we suspect that it is due to the Zener diode, which was likely not designed for low-temperature applications. 
In periods with low bias voltage, the PMT's QE is expected to be reduced to an unknown value.
To mitigate this in future runs, we plan to relocate the Zener diode to a room-temperature environment.

\subsection{The HeXe gas handling system}  
\label{subsec:HeXe}

A dedicated gas handling system, connected to the TPC, allows xenon storage, the introduction of gaseous calibration sources, and gas purification. During TPC operation, the xenon is continuously purified by circulating it through a hot getter. This is achieved by evaporating xenon from the liquid phase, circulating it as a gas, and then recondensing it on the cold head, from where it drips back into the TPC. Detailed information on the system is provided in~(\cite{Jorg:2021hzu}).

The xenon purity is assessed by measuring the ``electron lifetime'', a parameter that quantifies the loss of drifting electrons in the liquid due to attachment to electronegative impurities.
During this measurement campaign, the electron lifetime varied from about 40 to 50\,$\upmu$s.
This corresponds to an $\mathrm{O_2}$-equivalent mole fraction of approximately \SI{10}{ppb} using the electron attachment rate constant from~\cite{Bakale:1976_electron_attachment}.
The electron lifetime is notably shorter than in previous runs (a few hundred to a few thousand $\upmu$s), which was likely caused by outgassing from the IR PMT materials and the PTFE structure of the TPC.

\section{Calibration source, DAQ, and data selection} 
\label{sec:data}

The measurements were carried out using an internal $^{222}$Rn $\alpha$-particle calibration source.
This source was produced by depositing a 4.2\,g liquid sample with $\sim$29\,kBq of $^{226}$Ra onto a glass substrate.
The dried substrate was placed in a xenon-filled emanation chamber, where $^{222}$Rn accumulated over a period of one to three days. Subsequently, the radon-enriched xenon gas was expanded into a transfer chamber and introduced into the HeXe gas system. 
The resulting $\alpha$-particle rate in the TPC ranged from 20 to 30\,Hz, originating from the decays of $^{222}$Rn, $^{218}$Po, and $^{214}$Po, with energies of 5.5, 6.0, and 7.7 MeV, respectively.

Signals were triggered on the UV channel and recorded using a CAEN V1724 digitizer, which has a sampling rate of 100\,MS/s and 14-bit resolution.
Due to the nanosecond-scale rise and fall times of the IR signals, a timing filter amplifier was employed to shape the IR pulses, in order to maintain detection efficiency.
The digitized waveforms were processed to identify peaks, from which key characteristics -- such as arrival time, pulse area converted to the number of photoelectrons (PE), and width -- were extracted~(\cite{Cichon:2021phd}). Peaks in the UV channel were subsequently classified into S1 and S2 signals.

In addition to the nominal fields stated in \cref{subsec:tpc}, the TPC was also operated at different electric field strengths in the gas phase. 
Electric field values, including the drift and amplification fields above the gate, were determined and verified through COMSOL multiphysics simulations~(\cite{comsol}).
These simulations also provide estimates of the associated uncertainties for all field configurations.

\begin{figure}[t]
	\centering
	\includegraphics[width=\columnwidth]{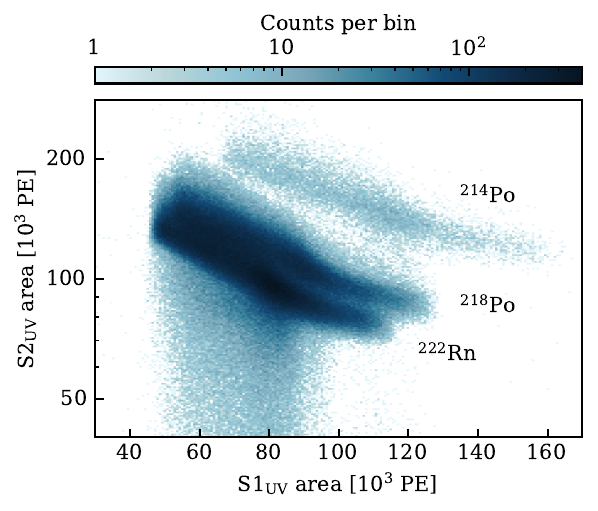}
	\caption{Distribution of $^{222}$Rn, $^{218}$Po and $^{214}$Po events in \soneuv-\stwouv area space after all cuts are applied.
    }
	\label{fig:ir_S1S2}
	\end{figure}

For IR pulses, only a peak height threshold is applied to suppress noise. A set of selection criteria is applied to the UV signals to select well-reconstructed $\alpha$-particle events.
Events within the liquid xenon target are selected based on their drift time.
Single scatter interactions are selected by requiring exactly one \soneuv and rejecting events with a secondary physical \stwouv following the primary.
To reject muon events and signals with distorted pulse shapes, we apply a cut on the \stwouv width and shape. Finally, $\alpha$-particle events are selected in the \soneuv-\stwouv parameter space, as shown in \cref{fig:ir_S1S2}.
The populations with approx. $70\times10^3$\,PE in \soneuv and $10^5$\,PE in \stwouv correspond to the three $\alpha$-decays of $^{222}$Rn. They have an extended distribution in light and charge signal due to position-dependent variations in light collection efficiency across the detector.

\section{Results}  
\label{sec:results}
Owing to the IR PMT’s small entrance window and a nominal quantum efficiency roughly five times lower than that of the UV PMT, its light collection efficiency is orders of magnitude lower.
Consequently, S1 and S2 signals of $\alpha$-particle events are recorded as thousands of photons in the UV PMT, but as individual photons in the IR PMT.

\begin{figure}[h]
	\centering
	\includegraphics[width=\columnwidth]{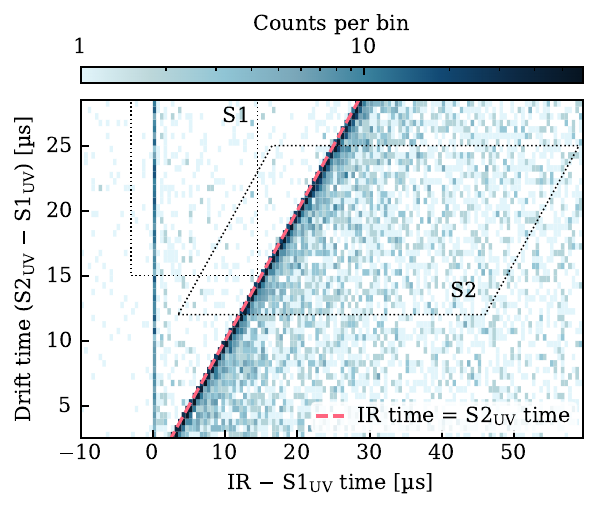}
	\caption{Time distribution of IR signals. The drift time of events with IR signals is plotted against the time difference between the IR signal and the \soneuv signal. A vertical band around zero corresponds to IR signals coinciding with \soneuv signals, while a diagonal feature represents signals coinciding with \stwouv signals, also indicated by the red dashed line. Regions selected to study the time response following the \soneuv (\cref{fig:ir_timing_s1}) and \stwouv (\cref{fig:ir_timing_s2}) are indicated with dotted lines.}
	\label{fig:ir_timeDiff}
\end{figure}

For each identified IR peak in a waveform, the time differences relative to the \soneuv and \stwouv signals of the event are calculated.
The pulse times for \soneuv and the IR channel are defined as the first sample preceding 10\% of the peak’s height.
For \stwouv signals, the time of maximum pulse height is taken.
\Cref{fig:ir_timeDiff} shows the drift time of the event (time between \soneuv and \stwouv signals) as a function of the time difference between IR pulses and the corresponding \soneuv.
Events at the top of the TPC have a short distance between \soneuv and \stwouv and are located at the bottom of \cref{fig:ir_timeDiff}. 

A small population of IR pulses are very close in time to the \soneuv across all drift times,  which we interpret as IR S1 signals. A larger population of IR pulses is found at or after the \stwouv time (to the right of the red dashed line). Already from this observation, we can conclude that we see IR emission associated with both \soneuv and \stwouv signals.

\subsection{Signal time structure}  
\label{subsec:Res_Timing}

\begin{figure}[h]
	\centering
    \includegraphics[width=\columnwidth]{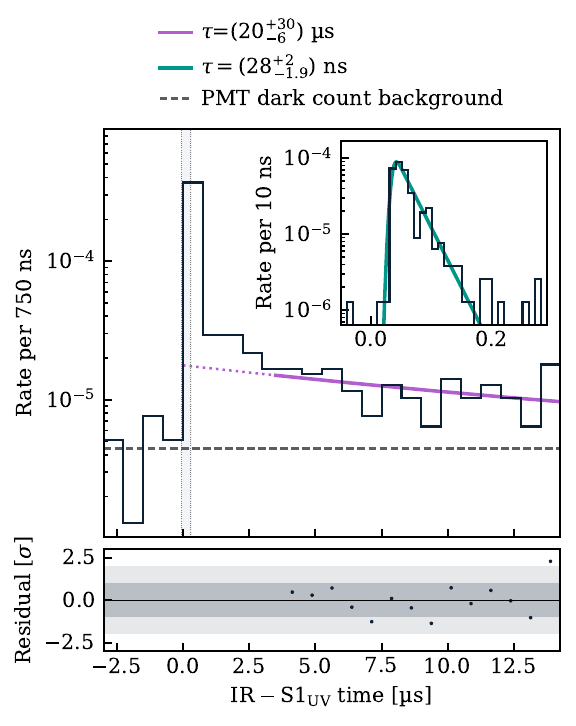}
    
	\caption{Time distribution of IR signals relative to the \soneuv time, with the TPC operated at a drift field of 400\,V/cm. The inset zooms into the region highlighted by the gray vertical band.  Fit results for the fast (green) and slow (purple) components are overlaid. Fits were performed using an unbinned likelihood; the binning and residuals shown are for visual reference only.}
	\label{fig:ir_timing_s1}
\end{figure}

We investigated the IR time response after \soneuv and \stwouv signals.
To study the time behavior following \soneuv, we selected signals before the \stwouv with drift times larger than \SI{15}{\upmu s} (dotted box labeled `S1' in \cref{fig:ir_timeDiff}), enabling the study of the distribution tail. \Cref{fig:ir_timing_s1} shows the arrival time distribution relative to the \soneuv up to 15$\upmu$s (main panel), with a magnification of the fast component in the inset up to 300\,ns.

The long IR tail following the \soneuv signal is well described by an exponential decay with a time constant of approximately $20 \upmu$s. The fast component, which accounts for roughly half of the total signal following the \soneuv, is modeled by an exponential with a time constant of $\sim$30\,ns, convolved with a Gaussian to account for the digitizer's timing resolution.
Although a precise determination of the fast time response is limited by the resolution and available statistics, we note that the observed scale is similar to previously reported values for \soneuv. For instance, \cite{Cichon:2022gbu} reports that approximately 65\% of the signal has a time constant of 26\,ns.

\begin{figure}[h]
	\centering
	\includegraphics[width=\columnwidth]{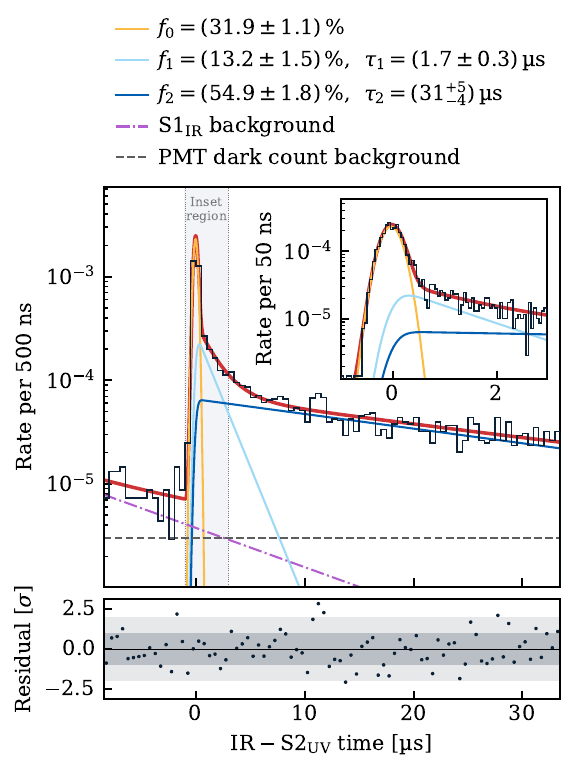}
	\caption{Time distribution of IR signals relative to the \stwouv time, measured at an electric field in the gas of 12\,kV/cm. The inset region is indicated as a gray vertical band. The total best-fit model (red) and its individual components are overlaid. Fits are performed using an unbinned likelihood; the binning and residuals shown are for visual reference only.}
	\label{fig:ir_timing_s2}
\end{figure}

To examine the IR emission associated with the electroluminescence process, we analyzed the arrival times of IR pulses occurring after the \stwouv (dotted box labeled `S2' in \cref{fig:ir_timeDiff}).
In \cref{fig:ir_timing_s2}, we show the time between each IR pulse and the \stwouv signal.
We model the distribution as the sum of three exponential decay components convolved with a Gaussian of $0.2\,\upmu$s width to account for the temporal dispersion of electron arrival times. An unbinned maximum likelihood fit is performed using this model. We include a flat background component to account for PMT dark count, as well as an exponential component originating from the IR signals following the \soneuv (see \cref{fig:ir_timing_s1}).

The fastest decay component is not well resolved due to the time smearing and is therefore simply modeled by a Gaussian. Its timescale may be comparable to the few-ns component previously observed in gaseous xenon at room temperature\,(\cite{Piotter:2023qli,Hammann:2024qyy}). The second fastest decay time of approx. $1.7\,\upmu$s is also similar to the decay time found in previous measurements. The third component of $\sim 30\,\upmu$s decay time has not been reported in xenon before.
This slowest component contributes more than 50\% of the total signal integral, making it the dominant contribution. Interestingly, this component is very similar to the slow component found after the \soneuv (purple line in \cref{fig:ir_timing_s1}), pointing to a common origin.
The fastest component accounts for approximately one-third of the total signal.

\begin{figure}[h]
	\centering
	\includegraphics[width=\columnwidth]{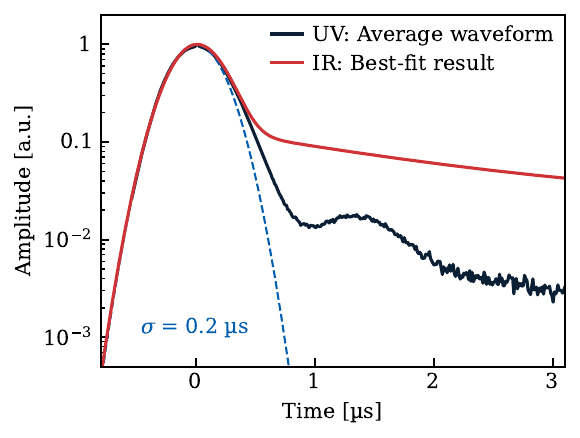}
	\caption{Average \stwouv pulse shape (black) overlaid with the background-subtracted best-fit IR time response following the \stwouv from  \cref{fig:ir_timing_s2} (red curve). For reference, a normal distribution with $\sigma=0.2\,\upmu$s is shown (blue dashed).}
	\label{fig:ir_S2ave}
	\end{figure}

\Cref{fig:ir_S2ave} compares the IR time response to the electroluminescence signal observed in the UV.
The average waveform of 500 \stwouv signals (black curve) is overlaid with the background-subtracted best-fit result from \cref{fig:ir_timing_s2}.
We find that the main \stwouv peak is well described by a Gaussian with a width of $\sigma = 0.2\,\upmu$s. The distribution has an asymmetric tail towards late times, which may result from PMT afterpulses and delayed electrons originating either from incomplete extraction or from photoionization of impurities in the liquid.
A small peak around 1.5\,$\upmu$s after the main signal is attributed to photoemission from the gate following the \stwouv, supported by its dependence on the amplification field and the observation of a similar delayed peak after large \soneuv signals.
While the main peaks of the UV and IR signals exhibit a similar shape, the IR signal features a significantly more pronounced tail -- approximately an order of magnitude larger in relative amplitude.
Moreover, the photoemission peak present in the UV signal is not observed in the IR, although it may be obscured by the extended tail.
The differences suggest that the origins of the respective tails are not the same.

\subsection{Electric field dependence}  
\label{subsec:E_dep}

To study the IR emission following \stwouv for different electron energies, we varied the amplification field in the xenon gas region from about 9\,kV/cm to 15\,kV/cm by varying the anode potential and keeping the drift field constant.
The IR PMT bias voltage was constant across all measurements.

An iterative procedure was employed to determine the liquid level.
Electrons generated by photoionization on the gate via bright UV signals were used, which have a drift distance corresponding to the liquid level.
Since the electron drift velocity depends on the electric field (\cite{Gushchin_1982_electron}), which itself is a function of the liquid level, the assumed level was adjusted iteratively until the calculated and observed drift times between the gate and the surface matched.
Averaging over all anode voltages and two analysis methods yielded a liquid level of $(4.1 \pm 0.5)$\,mm above the gate. The uncertainty is propagated to the electric fields in the liquid, which directly affect the extraction efficiency.
\Cref{tab:e_fields} summarizes the applied voltage differences, along with the calculated electric fields in the liquid and gas phases, based on the derived liquid level.
\begin{table}[h]
 \caption{Summary of potential differences between anode and gate and electric field values in the gas ($E_\mathrm{gas}$) and liquid phase ($E_\mathrm{liquid}$).
 } \label{tab:e_fields}
\begin{center}
\begin{tabular}{|c c c|} 
 \hline
 $\Delta$V [V] & $E_\mathrm{gas}$ [kV/cm] & $E_\mathrm{liquid}$ [kV/cm] \\
 \hline
 2\,775 & $8.9\pm0.7$  & $ 4.8\pm 0.4 $  \\
 \hline
 3\,775 & $12.1\pm0.9$  & $6.5\pm0.5
$  \\
 \hline
 4\,775 & $15.3\pm1.1$  & $8.3\pm0.6$  \\ 
 \hline
\end{tabular}
\end{center}
\end{table}

To count IR photons associated with the electroluminescence signal, we select those occurring within a time window from $-1$ to $34\,\upmu$s relative to the \stwouv, thereby minimizing contamination from signals related to the \soneuv.
Some of the selection criteria described in \cref{sec:data} may depend on the radial position of the event, which affects the light collection efficiency of the IR PMT and thus the observed IR photon rate.
To reduce possible bias, only minimal cuts are applied for this estimate, and the variation introduced by additional cuts is used to estimate a systematic uncertainty.
The observed IR photon rate is corrected for the field-dependent electron extraction efficiency (\cite{XENON100:2013wdu,PhysRevD.99.103024}). Potential systematic uncertainties on the extraction are taken into account. 

\begin{figure}[h]
	\centering
	\includegraphics[width=\columnwidth]{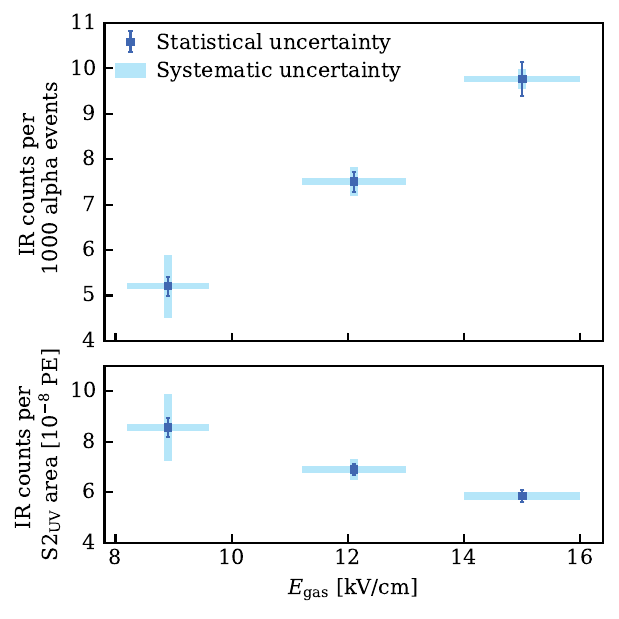}
	\caption{Top: Number of IR counts following the \stwouv per alpha events over the electric field in the gas phase. Bottom: Mean number of IR counts per \stwouv area.}
	\label{fig:ir_Efield}
\end{figure}

We observe a monotonic increase with the field strength in the IR photon rate, as shown in \cref{fig:ir_Efield} (top panel).
A similar trend was also observed in~\cite{Belogurov:2000} for lower fields.

While absolute yields cannot be determined, we compare the observed IR rate to the size of the associated \stwouv signal, as shown in the lower panel of \cref{fig:ir_Efield}.
The resulting ratio appears to deviate from a constant, suggesting that the two signals might not scale identically with the amplification field.
It is important to note that IR peaks with arrival times beyond 34\,$\upmu$s are not included in the rate estimates.
At the nominal amplification field of 12\,kV/cm, approximately 15\% of IR pulses fall outside the selected time window.
This fraction may vary with field strength due to changes in the slowest time constant, but limited statistics prevent a reliable estimate of this dependence.

\section{Discussion and outlook}  
\label{sec:discussion}

We investigated the scintillation response in a dual-phase xenon TPC in the wavelength range from 170\,nm to 1700\,nm. Infrared photons were observed in coincidence with UV scintillation from both the liquid and gas phases, and their time profiles were studied. As expected, increasing the electric field in the gas phase led to an increased number of IR photons associated to the electroluminescence signal.

The origin of the $\sim1\,\upmu\mathrm{s}$ IR scintillation constant in gas, observed here and also in our previous publication~(\cite{Hammann:2024qyy}), remains unclear. Since the UV scintillation decay constants are faster than the slow IR component, it is unlikely that this decay originates from a precursor state of the lower lying Xe$^{*}_2$ leading to the intense UV scintillation.
One possible explanation is a de-excitation pathway that bypasses the lowest-lying excimer state, thereby preventing the emission of UV light.

The IR-sensitive PMT used for the measurements presented here is also sensitive to shorter wavelengths as shown in \cref{fig:spectra_and_qe}, adding some uncertainty to the interpretation of the results.
In particular, it is sensitive to the tail of the third continuum emission, which has a decay constant of approximately 8\,ns\,(\cite{Millet:1978}).
This component has so far been measured as part of the primary scintillation light in gaseous xenon for $\alpha$-particles at pressures up to 10\,bar, as well as for electrons at 10\,bar~(\cite{Leardini:2021qnf}).
In electroluminescence, it was shown that the third continuum contributes at most at the sub-percent level to \stwouv.
Given the PMT’s roughly threefold lower QE at this wavelength and partial spectral overlap, its contribution is considered subdominant.
An additional contribution spanning a broad wavelength range arises from neutral bremsstrahlung~(\cite{henriques2022neutral,Leardini:2021qnf}).
To the best of our knowledge, neither of these processes has been systematically studied in the liquid phase.

In the IR time response following both \soneuv and \stwouv signals, we observed a time constant of approximately 30\,$\upmu$s, which accounts for roughly half of the respective total signals.
This similarity may point to a common origin of the slow component.
Notably, a comparable slow component has been observed in the primary IR scintillation of both liquid and gaseous argon (\cite{Buzulutskov:2011_nir_ar,Bondar:2012_nir_ar_part1}) with unresolved origin, though it may be attributed to a slow scintillation process.
Nevertheless, alternative explanations should also be considered.
Delayed electron tails have been observed in several liquid xenon TPCs (see for example~\cite{XENON100:2013wdu}).
A fraction of liberated electrons are not immediately extracted into the gas phase.
Instead, they are reflected off the liquid surface and eventually extracted at later times.
In addition -- in particular when the xenon purity is not particularly high -- UV scintillation light may photoionize residual impurities in the liquid, generating delayed electrons.
However, any signal from delayed electrons would also be visible in the UV channel, and thus can only account for the tail observed in the \stwouv signal.
Another possible contribution may arise from material fluorescence.
The TPC wall is made of PTFE, which has been suggested as a source of fluorescent light~(\cite{Shaw:07})), although it is debated~(\cite{Araujo_2019}) or attributed instead to fluorescence of residues in the PTFE~(\cite{Pollmann_2025}).
Fluorescence from quartz -- the material used for the UV PMT window -- has also been reported to produce a long signal tail in the UV-sensitive detectors~(\cite{Sorensen:2025_quartz}), but could also contribute to the slow component observed in the IR channel.
We also cannot rule out a contribution from a PMT-related effect, possibly induced by the suboptimal bias voltage applied to the photocathode, which may lead to delayed photoelectron extraction from the InP/InGaAs photocathode.

Although our data is not sensitive to the time constant of the fast IR scintillation component in the gas, it is sensitive to the amplitude of this component. Our past measurements at room temperature showed that, for $\alpha$-particles, the $\sim$\,ns component contributes less than 1\% to the total signal~(\cite{Hammann:2024qyy}). The current measurements with electrons in the gas show that the fast component makes up about one-third of the signal. This difference may result from variations in the population of excited states caused by differing ionization densities.
As a result, IR pulse shape discrimination could be a viable method for identifying particle types in the gas.

The measurements presented here cannot be used to quantify the IR light yield, primarily because the PMT’s quantum efficiency at the applied low bias voltage is unknown.
While it remains unclear whether IR scintillation in liquid xenon is sufficiently strong for practical applications, our data support the presence of IR emission from the electroluminescence process in xenon gas, consistent with previous observations.
Since previous measurements in gaseous xenon indicate a scintillation yield in the IR comparable to the UV, its application in particle detectors appears promising.
One potential application in future dual-phase liquid xenon detectors is the improved discrimination  of single-electron S2 signals from S1 signals.
Therefore, the ratio of IR and UV signals could eventually improve the rejection of accidental coincidences.
Beyond liquid detectors, IR scintillation in xenon gas could also offer advantages for high-pressure gaseous TPCs being developed to search for neutrinoless double beta decay.
In such experiments, the additional IR signal could contribute to improved energy resolution -- crucial for the experiment's sensitivity -- and may even enable further background suppression via the IR pulse shape, as mentioned above.

Future measurements will help to improve our understanding of the IR emission and enable us to evaluate its feasibility for future low-background experiments. However, new photosensor solutions will be necessary.
The current R5509-73 IR PMT is not suitable for low-background applications, as its intrinsic radioactivity ($\mathcal{O}(100$\,mBq) in U and Th) is too high for such rare event searches.

This first study shows signals in the IR-sensitive channel associated with both \soneuv and \stwouv, although important questions remain regarding their underlying mechanisms.
In the future, we plan to repeat the measurements using improved PMT electronics that support the correct bias voltage.
This is expected to eliminate potential distortions in the time response and enable reliable quantification of the light yield.
Moreover, we plan to operate the detector with higher xenon purity to investigate potential delayed signals caused by impurities, as well as possible distortions in the time response due to quenching effects.
We also aim to study the two signals independently: the fast component following \soneuv will be investigated using a faster digitizer, while the electroluminescence signal can be studied in a gas-only detector and extended to longer time scales to compare with known delayed electron signals. To further disentangle different contributions, we plan to introduce a wavelength filter in front of the IR PMT, isolating emission above 1\,$\upmu$m.

\section*{Conflict of Interest Statement}
The authors declare that the research was conducted in the absence of any commercial or financial relationships that could be construed as a potential conflict of interest.

\section*{Author Contributions}
RH: investigation, data curation, formal analysis, methodology, writing–original draft, visualization, writing–review and editing; KB: investigation, formal analysis and writing–review; SF: investigation, writing–review; LH: investigation, writing–review; TMU: concep\-tualization, project administration, supervision, writing–original draft, writing–review and editing.

\section*{Funding}
RH and KB acknowledge the financial support of the International Max Planck Research School for Precision Tests of Fundamental Symmetries (IMPRS-PTFS).
The project is supported by the Max Planck Society.

\section*{Acknowledgments}
We gratefully acknowledge the technical services of the Max-Planck-Institut f\"ur Kernphysik and thank Yannick Steinhauser for producing the rendered illustration of HeXe.
We also thank Edgar Sanchez for valuable discussions.

\bibliographystyle{Frontiers-Harvard}
\bibliography{refs}


\end{document}